# Cosmological models with negative constant deceleration parameter in Lyra geometry

[1,3] F.Rahaman, [1] N.Begum, [1] G.Bag and [2] B.C.Bhui

[1] Dept. of Mathematics, Jadavpur University,
Kolkata – 700032, India

[2] Dept. of Maths., Meghnad Saha Institute of Technology,
Kolkata-700150, India

[3] E.Mail: farook_rahaman@yahoo.com

## Abstract:

Bermann [ Nuovo Cimento B (1983), 74, 182 ] presented a law of variation of Hubble's parameter that yields constant deceleration parameter models of the Universe. In this paper, we study some cosmological models with negative constant deceleration parameter within the framework of Lyra geometry.

## Introduction:

The origin of structure in the Universe is one of the greatest cosmological mysteries even today. The present day observations indicate that the Universe at large scale is homogeneous and isotropy and the accelerating phase of the Universe (recently detected experimentally) [1]. It is well known that exact solutions of general theory of relativity for homogeneous space times belong to either Bianchi types or Kantowski-Sachs [2].

In last few decades there has been considerable interest in alternative theory of gravitation. The most important among them being scalar tensor theories proposed by Lyra [3] and Brans Dicke [3]. Lyra proposed a modification of Riemannian geometry by introducing a gauge function into the structure less manifold that bears a close resemblances to Weyl's geometry. In general relativity Einstein succeeded in geometrising gravitation by identifying the metric tensor with gravitational potentials. In scalar tensor theory of Brans-Dicke on the other hand, the scalar field remains alien to the geometry.

Lyra's geometry is more in keeping with the spirit of Einstein's principle of geometrisation since both the scalar and tensor fields have more or less intrinsic geometrical significance.

In consecutive investigations Sen [4] and Sen and Dunn [5] proposed a new scalar tensor theory of gravitation and constructed an analog of the Einstein field equation based on Lyra's geometry which in normal gauge may be written as

$$R_{ik} - \tfrac{1}{2} g_{ik} R + (3/2) *\phi_i *\phi_k - \tfrac{3}{4} g_{ik} *\phi_m *\phi^m = -8\pi G T_{ik} \qquad \ldots(1)$$

where $*\phi_i$ is the displacement vector and other symbols have their usual meaning as in Riemannian geometry.

Halford [6] has pointed out that the constant displacement field $*\phi_i$ in Lyra's geometry play the role of cosmological constant $\Lambda$ in the normal general relativistic treatment. According to Halford, the present theory predicts the same effects within observational limits, as far as the classical solar system tests are concerned, as well as tests based on the linearized form of field equations.

Subsequent investigations were done by several authors in scalar tensor theory and cosmology within the framework of Lyra geometry [6].

In this paper we are discussing homogeneous Bianchi-I and Kantowski-Sachs models with negative constant deceleration parameter within the framework of Lyra geometry. This study will be quite justified because recent experiments extensively show that the Universe is accelerating [1].

In section 2, we study Bianchi-I space-time in which the material distribution consists of a perfect fluid.

In section 3, we consider Kantowski-Sachs model in the presence of mass less scalar field with a flat potential based on Lyra geometry in normal gauge i.e. displacement vector

$$*\phi_i = (\beta(t),0,0,0) \qquad \ldots(2)$$

The paper ends with a summary in section 4.

## 2. Bianchi-I model:

We consider an axially symmetric Bianchi-I metric, which is taken as

$$ds^2 = -dt^2 + A^2(t) dx^2 + B^2(t) (dy^2 + dz^2) \qquad \ldots(3)$$

We take a perfect fluid form of energy momentum tensors

$$T_{ab} = (\rho + p) U_a U_b - p g_{ab} \; ; \; U_i U^i = 1 \qquad \ldots(4)$$

[ where $U^i$ is the four velocity, p is the pressure and $\rho$ is mass energy density]

We assume the equation of state as $p = m\rho \qquad [0 \leq m \leq 1] \qquad \ldots(5)$

Now choosing units such that $8\pi G = 1$, the field equation (1) for the metric (3) reduces to

$$2(A^1 B^1 / AB) + [(B^1)^2 / B^2] = \rho + \tfrac{3}{4}\beta^2 \qquad \ldots(6)$$

$$2(B^{11}/ B) + [(B^1)^2 / B^2] = -p - \tfrac{3}{4}\beta^2 \qquad \ldots(7)$$

$$(B^{11}/ B) + (A^{11} / A) + (A^1 B^1 / AB) = -p - \tfrac{3}{4}\beta^2 \qquad \ldots(8)$$

[ '1' denotes the differentiation w.r.t. 't' ]

We are going to consider only constant deceleration parameter model defined by

$$q = -[VV^{11} / (V^1)^2] = \text{constant} \qquad \ldots(9)$$

where $V = (AB^2)^{1/3}$ is the over all scale factor.

Here the constant is taken as negative (i.e. it is an accelerating model of the Universe)

The solution of eq.(9) is

$$V = [at + b]^{1/(1+q)} \qquad \ldots(10)$$

[ a , b are integration constants.]

This equation implies, the condition of expansion is $1 + q > 0$.

From eqs. (7) and (8), we get

$$(B^{11}/ B) - (A^{11}/ A) + [(B^1)^2 / B^2] - (A^1 B^1 / AB) = 0 \qquad \ldots(11)$$

Which can be integrated to give

$$B^2 A^1 - ABB^1 = h \qquad \ldots(12)$$

[ h is an integration constant.]

From eq. (10), we get

$$(A^1/A) + 2(B^1/B) = 3(V^1/V) = [3a / (1+q)][at + b]^{-1} \qquad \ldots(13)$$

From eq.(12), we get

$$(A^1/A) - (B^1/B) = h / V^3 = h[at + b]^{-3/(1+q)} \qquad \ldots(14)$$

Solving eqs.(13) and (14), we get

$$A = A_0[ at + b]^{\alpha} \exp [ 2\delta(at + b)^{\gamma} ] \qquad \ldots(15)$$

$$B = B_0[ at + b]^{\alpha} \exp [- \delta(at + b)^{\gamma} ] \qquad \ldots(16)$$

where $\alpha = [1/(q+1)]$, $\delta = [h(1 + q) / 3a( q - 2 )]$, $\gamma = [( q - 2 ) / ( 1 + q ) ]$

and $A_0$, $B_0$ are integration constants ]

**The physical quantities that are important in cosmology are proper volume $V^3$, expansion scalar $\theta$ and shear scalar $\sigma^2$ and Hubble's parameter H and have the following expressions for the above solutions :**

$$V^3 = [ at + b]^{3/(1+q)} \qquad \ldots(17)$$

$$\theta = [ 3a / ( 1+q) ] [ at + b]^{-1} \qquad \ldots(18)$$

$$\sigma^2 = (1/6)h^2 [ at + b]^{-6/(1+q)} \qquad \ldots(19)$$

$$H = [a / ( 1+q) ] [ at + b]^{-1} \qquad \ldots(20)$$

The expressions for $\beta^2$ (t) and $\rho$ are given by

$$\tfrac{3}{4} ( 1 - m )\beta^2 (t) = \tfrac{1}{3}ah\alpha ( 2m + 24 ) [ at + b]^{-(1+3\alpha)} - [a^2 \alpha ( 3m + 1)\alpha - 2][at + b]^{-2}$$

$$[\tfrac{1}{3}h^2(1 - m)][at + b]^{-6\alpha} \qquad \ldots(21)$$

$$( 1 - m ) \rho = [2a^2 \alpha ( 3\alpha - 1)][at + b]^{-2} + ( 4 ah\alpha /3) [at + b]^{-(1+3\alpha)} \qquad \ldots(22)$$

## 2.1: Behavior of the model:

We see that the Universe starts at an initial epoch $t_0 = - b / a$, which is a point singularity. At this instant, all the physical quantities diverge. Thus the Universe starts with an infinite rate of expansion and measure of anisotropy. So this is consistent with the big bang model. Also one can comment on the final stage of the evolution for this solution. As proper volume becomes infinitely large as $t \to \infty$, the other physical quantities such as density, pressure, shear e.t.c. become insignificants. Also one can note that shear tends to zero faster than the expansion.

In this model particle horizon exist because

$$\int_{t_0}^{t} dt^1 / V(t^1) = \{ (q+1) / aq \} \left[ \{at + b\}^{q/(1+q)} \right]_{t_0}^{t} \qquad \ldots(23)$$

is a convergent integral.

For this model, the gauge function was large in the beginning but decreases with the evolution of the model.

## 3. Kantowski-Sachs Model:

In this section we would like to consider Kantowski-Sachs model in the presence of mass less scalar field $\phi$ (t) with a flat potential $V(\phi)$.

The energy momentum tensor is taken as [6]

$$T_{ab} = \tfrac{1}{2} \partial_a\phi \, \partial_b\phi - [\, \tfrac{1}{4} (\partial_k\phi \, \partial^k\phi) + \tfrac{1}{2} V(\phi) \,] \, g_{ab} \qquad \ldots\ldots(24)$$

The $\phi$ field equation is [7]

$$(1/\sqrt{-g}) \, \partial_a (\sqrt{-g} \, \partial^a\phi) = - \, dV(\phi)/d\phi \qquad \ldots\ldots(25)$$

The metric ansatz for the Kantowski-Sachs space-time with topology of these spaces as $S^1 \times S^2$ is

$$ds^2 = dt^2 - A^2(t) \, dr^2 - B^2(t) \, (d\theta^2 + \sin^2\theta \, d\varphi^2) \qquad \ldots\ldots(26)$$

The field equation (1) reduces to ($\phi$ will be rescaled by $2\phi$)

$$2(A^1 B^1 / AB) + [(B^1)^2 / B^2] + (1/B^2) = (\phi^1)^2 + \tfrac{1}{2} V(\phi) + \tfrac{3}{4}\beta^2 \qquad \ldots\ldots(27)$$

$$2(B^{11}/B) + [(B^1)^2 / B^2] + (1/B^2) = -(\phi^1)^2 + \tfrac{1}{2} V(\phi) - \tfrac{3}{4}\beta^2 \qquad \ldots\ldots(28)$$

$$(B^{11}/B) + (A^{11}/A) + (A^1 B^1 / AB) = -(\phi^1)^2 + \tfrac{1}{2} V(\phi) - \tfrac{3}{4}\beta^2 \qquad \ldots\ldots(29)$$

$$\phi^{11} + \phi^1 [(A^1/A) + 2(B^1/B)] = (\phi^1)^2 \, [dV(\phi)/d\phi] \qquad \ldots\ldots(30)$$

[ The '$^1$' denotes the differentiation w.r.t. 't' ]

Here the potential can be approximated by a constant value {cf.Stein-Schabes,1987,[7]}

$$V(\phi) = 2\lambda \qquad \ldots\ldots(31)$$

It may be noted that the coefficient of $\phi^1$ in eq.(30) acts as a friction term and it is larger for an isotropic model. So the $\phi$ – field moves slowly in an anisotropic space-time.

Here we also consider constant deceleration parameter model.

Here $V = (AB^2)^{1/3} = [at + b]^{1/(1+q)}$ as above [ see eq.(10)].

Taking now the following combination of equations (27) – (29) + 2.(28), we get

$$(A^{11}/A) + 2(A^1 B^1 / AB) = \lambda \qquad \ldots\ldots(32)$$

From eq.(13), we get

$$(A^1/A) + 2(B^1/B) = 3(V^1/V) = [3a/(1+q)][at+b]^{-1} \qquad \ldots(33)$$

Eliminating $(B^1/B)$ from eq.(32) and eq.(33), we get

$$R^1 + R[3a/(1+q)][at+b]^{-1} = \lambda \qquad \ldots(34)$$

where $R = (A^1/A)$.

Solving eq.(34), we get

$$(A^1/A) = [\lambda(1+q)/a(4+q)][at+b] + D[at+b]^{-3/(1+q)} \qquad \ldots(35)$$

[ D is an integration constant.]

Thus the expressions of A and B are

$$A = \exp[\{\lambda(q+1)/2a^2(q+4)\}\{at+b\}^2 + \{D(q+1)/a(q-2)\}\{at+b\}^{(q-2)/(1+q)}] \qquad \ldots(36)$$

$$B^2 = [at+b]^{3/(1+q)} \cdot \exp - [\{\lambda(q+1)/2a^2(q+4)\}\{at+b\}^2 + \{D(q+1)/a(q-2)\}\{at+b\}^{(q-2)/(1+q)}] \qquad \ldots(37)$$

Using eq.(31), we get from eq.(30) as

$$\phi^{11} + \phi^1[(A^1/A) + 2(B^1/B)] = 0 \qquad \ldots(38)$$

Solving this we get, $\quad \phi^1 = [\phi_0 / A B^2] \qquad \ldots(39)$

( $\phi_0$ is an integration constant )

This implies

$$\phi = \int[\phi_0 / A B^2] dt + \phi_{00} \qquad \ldots(40)$$

[$\phi_{00}$ is another integration constant]

Thus we get an expression of $\phi$ as

$$\phi = [\{\phi_0(q+1)/a(q-2)\}\{at+b\}^{(q-2)/(1+q)}] + \phi_{00} \qquad \ldots(41)$$

Now the physical parameters take the following forms as

$$V^3 = [at + b]^{3/(1+q)} \qquad \ldots(42)$$

$$\theta = [3a/(1+q)][at+b]^{-1} \qquad \ldots(43)$$

$$\sigma^2 = (1/6)[\{3a/2(q+1)\}\{at+b\}^{-1} + \{(1+q)/2a(q+4)\}\{at+b\} + D\{at+b\}^{-3/(1+q)}] \quad ..(44)$$

$$H = [a/(1+q)][at+b]^{-1} \qquad \ldots(45)$$

$$\tfrac{3}{4}\beta^2 = [3\lambda/(4+q)] + 2\lambda + [\{3aD/(q+1)\}\{at+b\}^{-(q+4)/(1+q)}]$$

$$- [D^2 + \phi_0^2][\{at+b\}^{-6/(1+q)}] - [\{3a^2(q+4)/(q+1)^2\}\{at+b\}^{-2}$$

$$- [\{\lambda^2(q+1)/a^2(q+4)^2\}\{at+b\}^2 - [\{2\lambda D(q+1)/a(q+4)\}\{at+b\}^{(q-2)/(1+q)}]$$

$$- [at+b]^{-3/(1+q)} \cdot \exp[\{\lambda(q+1)/2a^2(q+4)\}\{at+b\}^2 + \{D(q+1)/a(q-2)\}\{at+b\}^{(q-2)/(1+q)}]$$

$$\ldots..(46)$$

## 3.1: Behavior of the model:

From the above solutions we note that at the initial epoch $t_0 = -b/a$, $A \to 1$ and $B \to 0$.
So it is a line singularity.
At the initial epoch $\theta$, $\sigma^2$ are infinitely large. We also see that at $t_0 = -b/a$, $\beta^2$ diverges. So the Universe starts from an initial singularity where all the physical parameters have infinite value, and then expand indefinitely. This model has particle horizon. As $t \to \infty$, the expansion ceases and the gauge function becomes imaginary. Thus concept of Lyra geometry will not linger for infinite time.

## 4: Summary:

In this work, we discussed two cosmological models namely Bianchi-I and Kantowski-Sachs models with constant deceleration parameter within the framework of Lyra geometry. In the first case, we have taken an ad hoc energy momentum tensor components but in the latter case we have used field theoretic approach with a flat potential.
Since our models consist with constant deceleration parameter, the proper volumes remain identical for both the models but the physical natures are different.
For both the models, Universe starts at an initial epoch $t_0 = -b/a$, and then expands indefinitely with an acceleration.

# Acknowledgements:

F.R is thankful to IUCAA for providing research facilities. We are also grateful to the referee for his valuable comments.